\documentclass[extra,referee]{gji}
\usepackage{etoolbox}

\usepackage{colin_gji}

{\title[Decadal zonal accelerations in Earth's core]
{Convectively driven decadal zonal accelerations in Earth's fluid core}}

\author[C. More and M. Dumberry]  
  {Colin More and Mathieu Dumberry \\    
  Department of Physics, University of Alberta, \\
  Edmonton, T6G 2E1, Canada, dumberry@ualberta.ca}

\begin{document}

\maketitle

\begin{abstract}

Azimuthal accelerations of cylindrical surfaces co-axial with the rotation axis have been inferred to exist in Earth's fluid core on the basis of magnetic field observations and changes in the length-of-day.  These accelerations have a typical timescale of decades.  However, the physical mechanism causing the accelerations is not well understood.  Scaling arguments suggest that the leading order torque averaged over cylindrical surfaces should arise from the Lorentz force.  Decadal fluctuations in the magnetic field inside the core, driven by convective flows, could then force decadal changes in the Lorentz torque and generate zonal accelerations.  We test this hypothesis by constructing a quasi-geostrophic model of magnetoconvection, with thermally-driven flows perturbing a steady, imposed background magnetic field.  We show that when the \alfven\ number in our model is similar to that in Earth's fluid core, temporal fluctuations in the torque balance are dominated by the Lorentz torque, with the latter generating mean zonal accelerations.  Our model reproduces both fast, free Alfv\'{e}n waves and slow, forced accelerations, with ratios of relative strength and relative timescale similar to those inferred for the Earth's core.  The temporal changes in the magnetic field which drive the time-varying Lorentz torque are produced by the underlying convective flows, shearing and advecting the magnetic field on a timescale associated with convective eddies.  Our results support the hypothesis that temporal changes in the magnetic field deep inside Earth's fluid core drive the observed decadal zonal accelerations of cylindrical surfaces through the Lorentz torque.

\end{abstract}

\begin{keywords}
core, numerical modelling, planetary interiors
\end{keywords}

\section{Introduction}
\label{sec:intro}

Decadal variations in Earth's length-of-day (LOD) are believed to be driven by core flow variations \citep{Gross2015}.  For instance, an increase in the bulk angular velocity of the core entrains an increase in its axial angular momentum.  Angular momentum conservation between the core and the mantle then implies that the latter must slow its rotation rate, thereby leading to a decrease in LOD.\\

\begin{figure}
    \centering
    \includegraphics[width=0.25\textwidth, keepaspectratio]{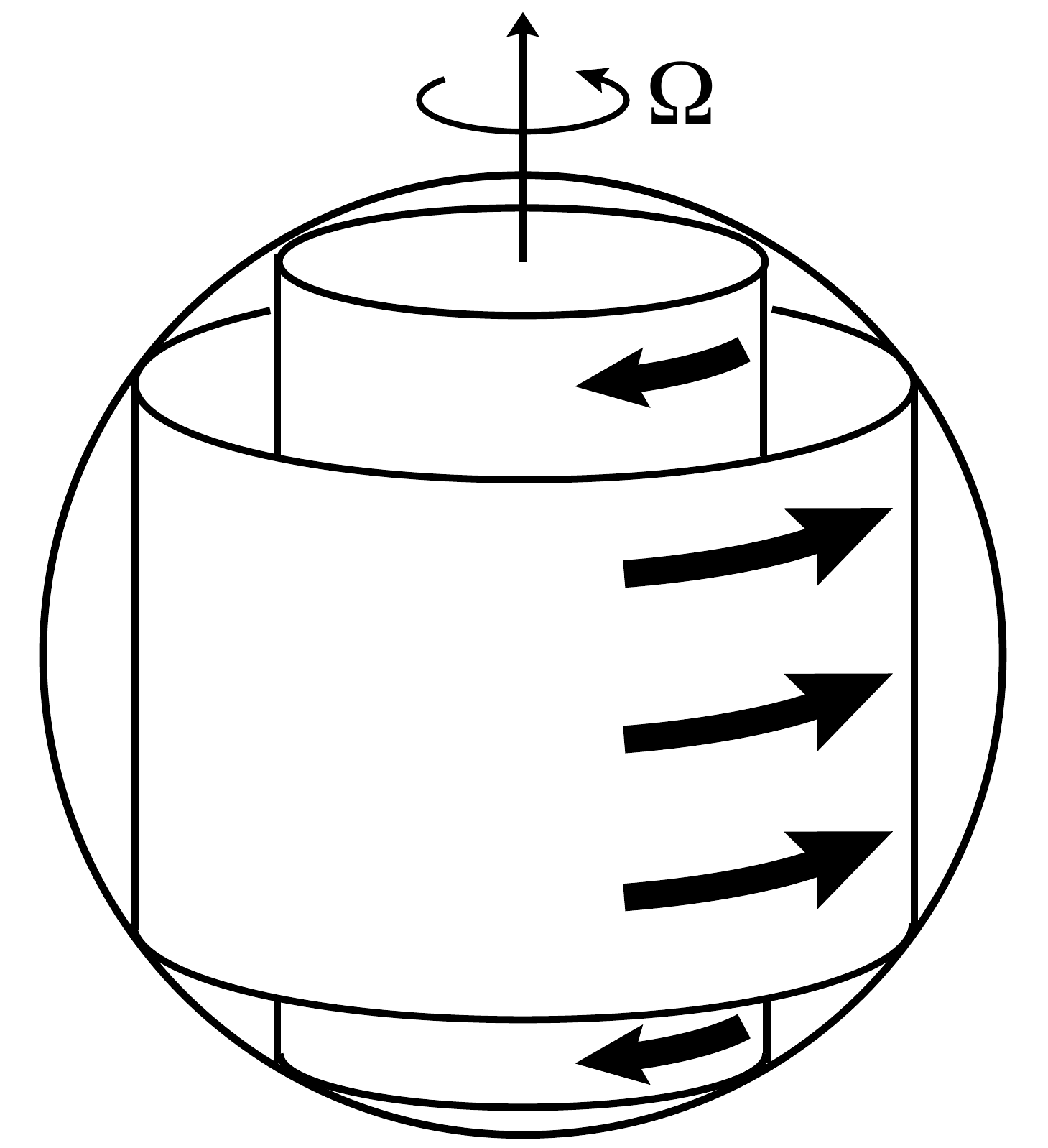}
    \caption{The geometry of the zonal flows in Earth's outer core which carry angular
      momentum.  The depicted time varying zonal flows consist in either free \alfven\ waves or forced accelerations.}
    \label{fig:geostrophic_cylinders}
\end{figure}

On timescales of centuries or less, the dynamics of the outer core are expected to be dominated by a geostrophic balance between pressure gradients and the Coriolis force \citep[e.g.][]{Finlay2010}. Flows that obey such a balance have the property of being invariant, or ``rigid'', parallel to the axis of rotation \citep{hough1897,proudman1916,taylor1917}.  This is the Taylor-Proudman theorem. Temporal variations in the outer core's axial angular momentum must then be carried by azimuthal accelerations in the form of rigid, co-axial cylindrical surfaces, or ``geostrophic cylinders'', as shown in \Figure \ref{fig:geostrophic_cylinders}.\\

Flows at the core-mantle boundary (CMB) may be reconstructed from the magnetic field's secular variation observed at Earth's surface \citep{Holme2015}.  Time-dependent accelerations of geostrophic cylinders may then be extracted from the reconstructed core flows, allowing a prediction of LOD changes to be built.  A number of studies \citep[e.g.][]{Jault1988, Jackson1993} have shown that such predictions match well with the observed decadal LOD variations.  Furthermore, \citet{gillet2010} have shown that zonal accelerations of geostrophic cylinders carry changes in angular momentum which can also explain an observed 6-yr periodic LOD signal \citep{Holme2013,Chao2014}.\\

These studies confirm not only that the decadal and 6-yr LOD variations are due to core-mantle angular momentum exchanges, they also confirm the presence of rigid zonal accelerations in the core. However, the dynamics responsible for such zonal accelerations are not fully understood.  This is the topic of our study.\\

The forces responsible for controlling the dynamics of zonal flows can be examined by integrating the azimuthal component of the momentum equation over the surface of geostrophic cylinders.  Upon doing so, the pressure, Coriolis, and buoyancy terms vanish.  If one neglects inertia and viscosity, this implies that the azimuthal component of the Lorentz force must cancel out when integrated over geostrophic cylinders. In other words, the axial Lorentz torque on cylinders must vanish.  This is known as Taylor's condition \citep{Taylor1963}, with systems obeying it said to be in a Taylor state.  Reinstating inertia, perturbations in the magnetic field -- and, therefore, in the Lorentz torque -- can be accommodated by rigid zonal accelerations of geostrophic cylinders.\\

When a magnetic field in a conducting fluid is distorted by the fluid's motion, such as the differential rotation between coaxial cylinders, a current is induced.  This current interacts with the original magnetic field to create a restoring force which opposes the motion that caused the initial magnetic distortion.  In the language of Taylor's condition, this restoring force nudges geostrophic cylinders back towards a Taylor state, subject to magnetic diffusion.  This mechanism allows rigid zonal flows to oscillate about a Taylor state and, in doing so, support \alfven\ waves.\\

\citet{Braginsky1970} suggested that the free modes of \alfven\ waves could be responsible for decadal zonal accelerations in the outer core.  In order to produce free \alfven\ waves with a fundamental mode period of approximately 60 years, corresponding to the characteristic timescale of the LOD signal \citep{Roberts2007,Gross2015}, the magnetic field strength throughout the core must be approximately 0.3 mT, similar to its observed strength at the CMB.\\

However, modern geodynamo simulations suggest that the internal radial magnetic field is approximately ten times larger than that at the CMB \citep[e.g.][]{christensen2006a}.  Such a field strength should yield a fundamental \alfven\ period of approximately 6 years.  It is therefore now believed that the 6-yr LOD signal is due to the propagation of free \alfven\ waves \citep{gillet2010}, leaving the dynamics responsible for the decadal zonal accelerations unexplained.\\

One possible explanation is that convective eddies in the fluid core continuously distort the magnetic field, causing spatial and temporal variations in the Lorentz force. Integrated over geostrophic cylinders, the induced Lorentz torques must be balanced by rigid zonal accelerations of the cylinders.  Continual distortion of the internal magnetic field would then lead to continual zonal accelerations of the cylinders.  The observed decadal zonal accelerations could therefore be a forced fluctuation about the Taylor state driven by convective flows.\\

If this is the case, forcing must occur on timescales longer than the propagation time of free \alfven\ waves.  In other words, the typical convective velocity $u_{C}$ must be smaller than the \alfven\ wave velocity $u_{A}$.  The relationship between $u_{C}$ and $u_{A}$ is captured by the \alfven\ number

\begin{equation}
    \alfnum = \frac{u_{C}}{u_{A}},
    \label{eq:alfven_number}
\end{equation}

where

\begin{equation}
    u_{A} = \frac{\abs{\bld{B}}}{\sqrt{\rho \mu_{0}}},
    \label{eq:alfven_velocity}
\end{equation}

$\bld{B}$ is the magnetic field, $\rho$ is the fluid density, and $\mu_{0}$ is the magnetic permeability of free space.  Typical large-scale flow velocities in Earth's core are of the order of $u_{C} = 10$ km yr$^{-1}$, or $3 \cdot 10^{-4}$ m s$^{-1}$ \citep{Holme2015}.  With a typical radial magnetic field of 3 mT \citep[eg.][]{gillet2010,buffett2010}, an outer core density of $10^{4}$ kg m$^{-3}$, and $\mu_{0} = 4 \pi \cdot 10^{-7}$ N A$^{-2}$, the \alfven\ wave velocity is approximately $u_{A} = 3 \cdot 10^{-2}$ m s$^{-1}$.  This yields $\alfnum \approx 0.01$.  Such a small value of $\alfnum$ supports the idea that the observed decadal zonal flows in Earth's core may be driven by convection via time-dependent Lorentz torques.\\

The goal of this study is to demonstrate that such a dynamical scenario is possible.  One option to investigate the zonal flow dynamics in the Earth's core is to use a 3D numerical model of a self-generated dynamo \citep[e.g.][]{Christensen2015}.  However, typical \alfven\ numbers in such models are $\alfnum \approx 1$, meaning the dynamics of free \alfven\ waves and convectively-driven zonal accelerations are not well separated.  Some recent 3D models have been able to achieve lower values of $\alfnum \approx 0.1$ \citep[e.g.][]{Aubert2017, Schaeffer+2017}, though these models are numerically expensive to run.\\

Here, we follow a different strategy and exploit the fact that, as discussed previously, the large-scale fluid motions in the core which vary on decadal timescales are expected to be almost invariant along the rotation axis \citep{jault08}.  Flows of this type are often termed Quasi-Geostrophic (QG), with their existence in Earth's core supported by the observed geomagnetic secular variation \citep{pais08,gillet11}.  They also emerge from numerical models of the geodynamo \citep[e.g.][]{Schaeffer+2017}.\\

We present in this study a numerical model of the decadal timescale dynamics of Earth's fluid core constructed within a QG framework \citep[e.g.][]{Aubert2003}.  We add an induction equation to a QG model of thermal convection to follow the evolution of the magnetic field as it is sheared and advected by the flow.  We likewise take into account the feedback of the Lorentz force on the flow, though in a limited way -- see Section \ref{sec:model:parameters}.  Since we focus on the short timescale dynamics, we substitute the self-generated magnetic field resulting from dynamo action with a steady, imposed background magnetic field. We thus perform a magnetoconvection experiment, in which the perturbations of the magnetic field are tracked with respect to this imposed field.  This strategy allows us to readily achieve $\alfnum < 1$ at very modest numerical cost.\\

\section{Model}
\label{sec:model}

\subsection{Background}
\label{sec:model:background}

\begin{figure}
    \centering
    \includegraphics[width=0.5\textwidth, keepaspectratio]{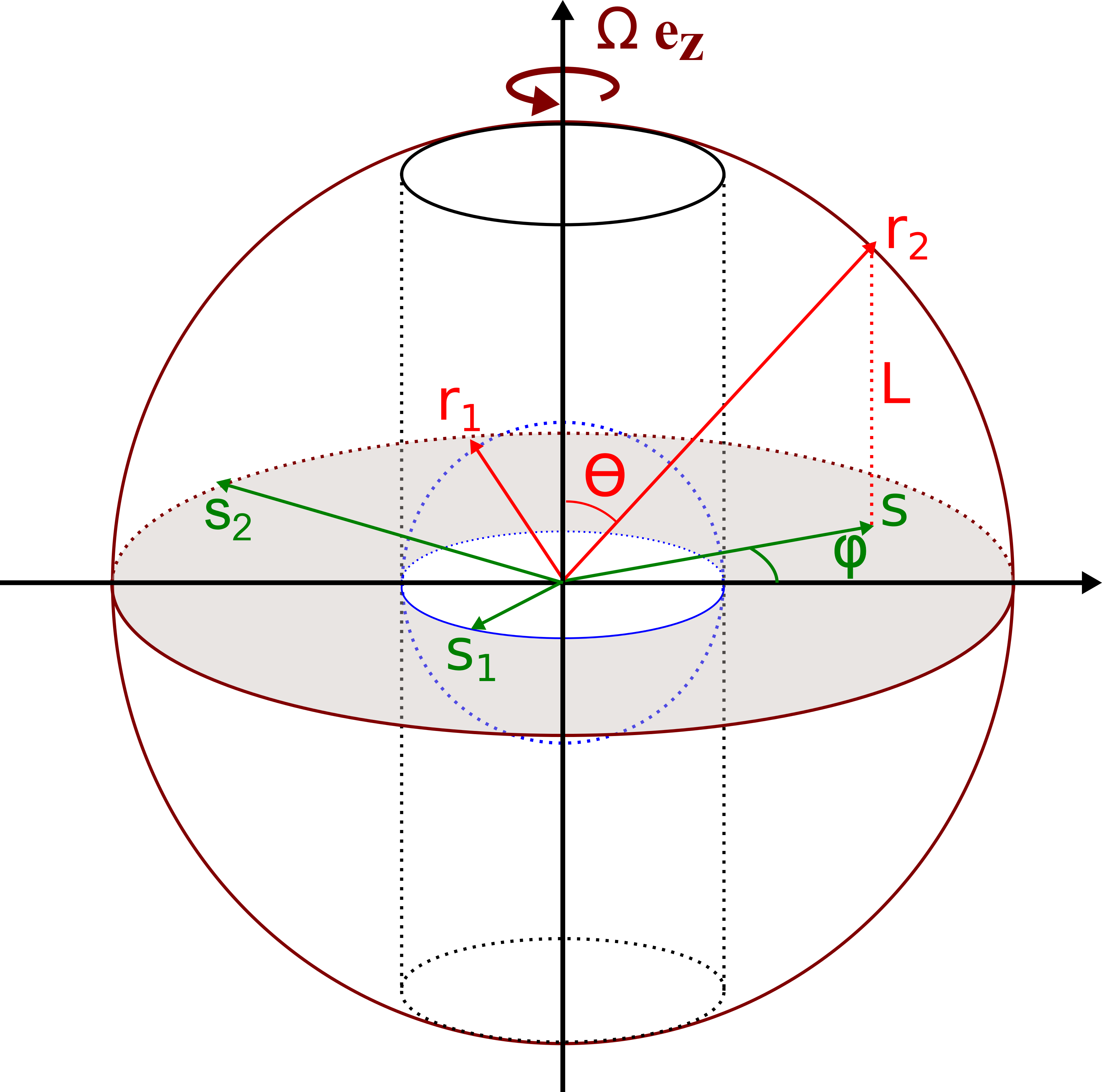}
    \caption{Geometry of our QG model.  The domain of integration is the shaded 2D annulus
      between $s_{1}$ and $s_{2}$.}
    \label{fig:qg_geometry}
\end{figure}

Previous studies \citep[e.g.][]{Aubert2003,Gillet2006} have shown that thermally-driven, Boussinesq, QG models can reproduce flow patterns similar in scale and behaviour to the ones we expect in planetary cores.  The QG approximation takes advantage of the geometric constraint imposed by strong rotation: fluid motions must be dominantly rigid.  Studying the dynamics of the full system is then equivalent to studying the dynamics of a slice through the system's equatorial plane.  A QG model therefore collapses three-dimensional convection experiments onto a two-dimensional domain.  In our case, this domain corresponds to the shaded annulus of Fig.\ref{fig:qg_geometry}.  Using the usual cylindrical coordinates $(s,\phi, z)$, with the \ez\ direction aligned with the rotation axis, this annulus extends from the ``tangent cylinder'' surrounding the equivalent of the inner core at $s=s_{1}$, to the ``equator'' at $s=s_{2}$.  We do not model the region inside the tangent cylinder, equivalent to the polar regions above and below the inner core.\\

Our model is based on the QG model of thermal convection, first presented by \cite{busse1986a} and \cite{Cardin1994} and expanded in many subsequent studies \citep[e.g.][]{Aubert2003,Gillet2006}.  In this model, two-dimensional equations for the momentum and thermal evolution are coupled together.  To these governing equations, we add a third, two-dimension magnetic induction equation, while augmenting the momentum equation with a Lorentz force term.  Although the Taylor-Proudman theorem predicts rigid motions, there are no such theoretical constraints on the temperature nor the magnetic field.  However, by averaging the governing equations in the axial (\ez) direction, we capture, if imperfectly, the net effect of both the temperature and magnetic field on the flow dynamics. Strategies have been devised to couple QG flows to a three-dimensional magnetic field \citep{schaeffer06,schaeffer16} and temperature \citep{guervilly16}, but here we restrict our attention to a purely two-dimensional model.\\

We scale length by the radius of the outer sphere $r_{2}$, time by the inverse of the angular rotational velocity $\Omega$, temperature by the superadiabatic temperature difference $\Delta T$ between the inner and outer spheres, and the magnetic field by $r_{2} \Omega \sqrt{\rho_{0} \mu_{0}}$, where $\rho_{0}$ is the reference density.  The QG approximation remains valid so long as the Coriolis force remains dominant.  In our case, this implies that both the Rossby number $R_{o}$ (the ratio of inertial to Coriolis forces) and the Lehnert number $\lambda$ (the ratio of magnetic to Coriolis forces) must remain $\ll 1$. Our choice of scalings causes $R_{o}$ in our model to be equivalent to $\abs{\uu}$ (the typical amplitude of the non-dimensional velocity) and $\lambda$ to be equivalent to $\abs{\bb}$ (the typical amplitude of the non-dimensional magnetic field perturbation).  We must therefore ensure that $\abs{\uu} \ll 1$ and $\abs{\bb} \ll 1$.

\subsection{Flow and Magnetic Fields}
\label{sec:model:fields}

Because both the flow and magnetic fields are solenoidal, they may be represented in terms of vector potentials.  We define the velocity $\uu$ and magnetic field perturbation $\bb$ as

\begin{subequations}
\begin{align}
    \uu & = \axiup \, \ep + \frac{1}{L} \del \times \pl L \psi \, \ez \pr \, , \\
    \bb & = \axibp \, \ep + \frac{1}{L} \del \times \pl L a \, \ez \pr \, ,
\end{align}
    \label{eq:full_field_definitions}
\end{subequations}
where $\psi$ and $a$ are toroidal scalars and $L = \sqrt{1 - s^{2}}$ is the half-column height. With an overbar denoting an azimuthal average, \axiup\ and \axibp\ capture the axisymmetric azimuthal (zonal) flow and zonal magnetic fields, respectively.  The horizontal components of the velocity field $\bld{u}_{H} = \pus \es + \pup \ep$ and the axial vorticity $\pwz$ are then defined as

\begin{subequations}
\begin{align}
    \pus & = \frac{1}{s} \pdphia{\psi} \, ,\\
        \pup & = \axiup - \pl \pds + \beta \pr \psi \, , \\
            \pwz & =   \pl 2 \vorO + \pds \axiup \pr
           - \n_{H}^{2} \psi
           - \frac{1}{s} \pds \pl s \beta \psi \pr \, .
\end{align}
    \label{eq:velocity_field_definition}
\end{subequations}

Similarly, the horizontal components of the magnetic perturbation field $\bld{b}_{H} = \pbs \es + \pbp \ep$ and the axial current $\pjz$ are defined as

\begin{subequations}
\begin{align}
    \pbs &= \frac{1}{s} \pdphia{a} \, , \\
        \pbp & = \axibp - \pl \pds + \beta \pr a \, , \\
    \pjz &=   \pl 2 \frac{\axibp}{s} + \pds \axibp \pr
           - \n_{H}^{2} a
           - \frac{1}{s} \pds \pl s \beta a \pr \, .
\end{align}
    \label{eq:magnetic_field_definition}
\end{subequations}

Here, $\n_{H}^{2}$ specifies the cylindrical $(s,\phi)$ components of the Laplacian operator.  The factor $\beta$, which enters the equations via the no-penetration condition on the $r=r_{2}$ surface, is a measure of how $L$ changes with $s$:

\begin{equation}
    \beta = \frac{1}{L} \pdsa{L} = - \frac{s}{L^{2}} \, .
    \label{eq:beta_definition}
\end{equation}

The definitions of \pus, \pup, and \pwz\ according to \Equations{\ref{eq:velocity_field_definition}} follow the traditional approach in QG models \citep[e.g.][]{Schaeffer2005}.  Because the fluid is assumed incompressible, there is no axisymmetric radial velocity.  Using an equivalent representation for \pbs, \pbp, and \pjz\ in \Equations{\ref{eq:magnetic_field_definition}} follows the strategy employed by \cite{Labbe2015}.  These rigid variables represent the axial averages of the truly three-dimensional magnetic field.  Since their form is equivalent to the flow field, the implied assumption is that the magnetic field obeys the equivalent of a no-penetration condition on the spherical top and bottom boundary of a fluid column.  While this cannot be rigorously justified, to first order this is approximately correct since the magnetic field near the core-mantle boundary is likely much smaller than deeper in the core.  Furthermore, as we show in Appendix \ref{appendix:angular_momentum}, representing the magnetic field as in \Equations{\ref{eq:magnetic_field_definition}} is necessary to ensure conservation of angular momentum.

\subsection{Momentum Equation}
\label{sec:model:momentum}

Under the QG approximation, the axially-averaged, thermally-driven Navier-Stokes equation reduces to an axial vorticity equation \cite[e.g.][]{Aubert2003},

\begin{equation}
    \pdt \pwz = - \Rastar \pdphia{\Theta}
                + E \n_{H}^{2} \pwz
                + \pl 2 + \pwz \pr \beta \pus
                - \pl \pus \pds + \frac{\pup}{s} \pdphi \pr \pwz
                + F_{L} \, .
    \label{eq:vorz_nonaxi}
\end{equation}

Here, $\Rastar = E^{2} \Ra P_{r}^{-1}$ is the modified Rayleigh number, $E = \nu \pl \Omega r_{2}^{2} \pr^{-1}$ is the Ekman number, $\Ra = \alpha g_{0} \Delta T r_{2}^{3} \pl \nu \kappa \pr^{-1}$ is the Rayleigh number, $P_{r} = \nu \kappa^{-1}$ is the Prandtl number, and $\nu$, $\alpha$, $g_{0}$, and $\kappa$ are respectively the kinematic viscosity, thermal expansion coefficient, gravitational acceleration at $r=r_{2}$, and thermal diffusivity.  $\Theta$ is the local temperature perturbation from the conducting profile (see Section \ref{sec:model:thermal}), and $F_{L}$ represents the axial component of the curl of the Lorentz force.\\

Since we wish to study the dynamics of zonal flows, we write a separate equation for the evolution of the axisymmetric angular velocity \vorO:

\begin{equation}
    \pdt \pl \vorO \pr = \gammaL + \gammaR + \gammaV \, ,
    \label{eq:vorz_axi}
\end{equation}

where \gammaL\ denotes the axial torque from Lorentz forces, \gammaR\ the axial torque from Reynolds stresses, and \gammaV\ the axial viscous torque.  \gammaR\ and \gammaV\ are given by

\begin{align}
    \label{eq:reynolds_axi}
    \gammaR &=        -  \frac{1}{s} \pl \moyp{\frac{\pus}{s} \pds s \pup} \pr, \\
    \label{eq:viscous_axi}
    \gammaV &=\phantom{-} \frac{E}{s^{3} L} \pds \pl s^{3} L \pds \pl \vorO \pr \pr,
\end{align}

while \gammaL\ is given in the next section by \Equation{\ref{eq:lorentz_axi_main}}.  Although \Equation{\ref{eq:vorz_axi}} is contained in \Equation{\ref{eq:vorz_nonaxi}}, in practice we use \Equation{\ref{eq:vorz_nonaxi}} to evolve the non-axisymmetric part of \pwz, and \Equation{\ref{eq:vorz_axi}} to evolve \axiup.\\

We note that we have neglected in both Eqs.~(\ref{eq:vorz_nonaxi}) and (\ref{eq:vorz_axi}) viscous friction at the top and bottom spherical boundaries.  These can be incorporated in the QG framework through Ekman pumping terms \citep[e.g.][]{Schaeffer2005}. However, for the Ekman numbers that we can reach numerically, these friction terms would play an unrealistically dominant role in the force balance, unlike in the Earth's core.

\subsection{Induction Equation and Lorentz Force}
\label{sec:model:magnetic}

Since a two-dimensional model cannot maintain a self-sustaining dynamo \citep[e.g.][]{Cowling1957,Roberts2015}, we instead impose a steady background magnetic field $\bld{B_{0}}$ on it.  For simplicity we choose a uniform field, $\bld{B_{0}}(s,\phi,z) = \backbs \, \es$.  This uniform field can be interpreted to represent the local cumulative effect from all length scales of the background field.  Two-dimensional perturbations $(\pbs, \pbp)$ from that background state are then tracked via the induction equation.\\

We form an equation for the magnetic potential $a$ by axially averaging the $s$-component of the induction equation, which yields

\begin{equation}
  \pdt a = - \pup \backbs
           + \pl \pus \pbp - \pup \pbs \pr
           + \frac{E}{\Pm} \pl \n_{H}^{2} a + \frac{2 \beta a}{s} \pr \, ,
    \label{eq:induction_nonaxi}
\end{equation}
where $\Pm = \nu \eta^{-1}$ is the magnetic Prandtl number and $\eta$ is the magnetic diffusivity. The $\beta$ factor in the last term of Eq. (\ref{eq:induction_nonaxi}) originates from the contribution of $b_{\phi}$ to the vector Laplacian in the $s$-direction.  Physically, it represents an added contribution to dissipation introduced by the spherical geometry.  \\

The equation for $\overline{b_\phi}$ is derived from the axially-averaged and azimuthally-averaged $\phi$-component of the induction equation,

\begin{equation}
    \pdt \pl \curO \pr = \backbs \pds \pl \vorO \pr
                     + \frac{1}{s}\frac{1}{L} \pds \Big( L \pl  \moyp{\pup \pbs} - \moyp{\pus \pbp} \pr \Big)
                     + \frac{1}{s^{3}} \frac{E}{\Pm} \pds \pl s^{3} \pds \pr \pl \curO \pr \, .
    \label{eq:induction_axi}
\end{equation}

The axially-averaged Lorentz force which enters \Equation{\ref{eq:vorz_nonaxi}} is

\begin{equation}
    F_{L} = \pl \pl \backbs + \pbs  \pr \pds + \frac{\pbp}{s} \pdphi \pr \pjz \, .
    \label{eq:lorentz_nonaxi}
\end{equation}

Although \Equation{\ref{eq:lorentz_nonaxi}} is the correct expression for the Lorentz force acting on flow eddies in our QG model, we note that in practice we do not compute this term, i.e. we set $F_{L}=0$. The justification for this is given in section \ref{sec:model:parameters}.  Meanwhile, the axial Lorentz torque in \Equation{\ref{eq:vorz_axi}} is

\begin{equation}
    \gammaL = \gammaLone + \gammaLtwo \, ,
    \label{eq:lorentz_axi_main}
\end{equation}

where

\begin{align}
    \gammaLone & = \frac{1}{s^{3}} \frac{1}{L} \pds \pl s^{3} L \backbs \curO \pr \, ,
    \label{eq:lorentz_axi_1} \\
    \gammaLtwo & = \frac{1}{s} \pl \moyp{\frac{\pbs}{s} \pds s \pbp} \pr \, .
    \label{eq:lorentz_axi_2}
\end{align}

\subsection{Thermal Equations}
\label{sec:model:thermal}

Thermal convection is driven by the steady, superadiabatic temperature difference $\Delta T = T_{1} - T_{2}$, where $T_{1}$ and $T_{2}$ are the superadiabatic temperatures on the spheres $r=r_{1}$ and $r=r_{2}$, respectively.  Taking the axial average of the resulting three-dimensional temperature profile $T_{0}$ gives the background conducting, rigid temperature profile $T$ as a function of cylindrical radius:

\begin{equation}
    T = \pl \frac{r_{2} T_{2} - r_{1} T_{1}}{r_{2} - r_{1}} \pr
      + \pl \frac{r_{1} r_{2}}{r_{2} - r_{1}} \pr \frac{\Delta T}{L} \ln \pl 1 + L \pr \, .
    \label{eq:background_temperature_profile}
\end{equation}

Since it is implicitly contained in the control parameter \Rastar, we are free to set $\Delta T$ to an arbitrary nonzero value.  We therefore choose $T_{1} = 1$, $T_{2}$ = 0.  These choices, along with \Equation{\ref{eq:background_temperature_profile}}, are consistent with those used by \cite{Aubert2003} and \cite{Gillet2006}.\\

Time-dependent perturbations $\Theta$ from $T$ are tracked with the axially-averaged heat equation

\begin{equation}
    \pdta{\Theta} = -\pl \uu_{H} \cdot \del_{H} \pr \pl T + \Theta \pr + \frac{E}{P_{r}} \n^{2}_{H} \Theta \, ,
    \label{eq:heat}
\end{equation}
where $\del_{H}$ represents the $(s, \phi)$ components of the gradient operator.

\subsection{Boundary Conditions}
\label{sec:model:boundary_conditions}

We ensure that the magnetic perturbation field \bb\ drops to zero at the boundaries by imposing $a = 0$, and $\curO = 0$.  Similarly, we ensure that the temperature perturbation drops to zero on the boundary by imposing $\Theta = 0$.  To respect the no-penetration boundary condition, $\psi$ must be constant on the inner and outer boundaries.  For convenience, we use $\psi = 0$.  We apply an additional no-slip boundary condition on the non-axisymmetric part of \pup\ such that $\pdsa{\psi} = 0$.  Forcing $u_s$ and $u_\phi$ to be both zero at the boundary is self-consistent with the $a=0$ condition (see Eq.~\ref{eq:induction_nonaxi}) so it is the most natural choice.  However, we impose a free-slip boundary condition on $\vorO$, such that $\pds\pl\vorO\pr=0$.  The latter choice is made to minimize viscous friction effects on the zonal flows -- the target of our study -- which should be small in an Earth-like regime.

\subsection{Parameter Regime}
\label{sec:model:parameters}

As argued in the introduction, in order to be in an Earth-like regime ($\alfnum \ll 1$), typical convective flow speeds ($u_{C}$) must be much smaller than the typical \alfven\ wave speed (\ualf). For a given Ekman number $E$, convective speeds are controlled by \Rastar.  With our choice of non-dimensionalization, the \alfven\ wave velocity in \Equation{\ref{eq:alfven_velocity}} is

\begin{equation}
    \ualf = \abs{\bld{B}} = \abs{\bld{B}_{0} + \bb} \, ,
    \label{eq:va_ndim}
\end{equation}

so a lower bound is set by the strength of the background magnetic field \backbs.  Therefore, the combination of \Rastar\ and \backbs\ in our model must be such that $\alfnum \ll 1$.\\

Furthermore, we must be in a regime where accelerations of the zonal flow are dominantly controlled by the Lorentz, rather than the Reynolds, torque.  An inspection of Eqs. (\ref{eq:reynolds_axi}) and  \ref{eq:lorentz_axi_2}) shows that the Reynolds and Lorentz torques are proportional to the nonaxisymmetric parts of $u^{2}$ and $b^{2}$, respectively, with typical scales given by their root-mean-square amplitudes $\urms = \abs{\uu - \moyp{\pup} \ep} \approx u_{C}$ and $\brms = \abs{\bb - \moyp{\pbp} \ep}$.  Hence, we must be in a regime where $\brms \gg \urms$.\\

The parameters controlling the ratio between $\brms$ and $\urms$ can be established by analyzing the induction equation.  When our model has achieved statistical equilibrium, the source and diffusion terms must be in balance:

\begin{equation}
    \abs{\del \times \pl \pl \bld{B}_{0} + \bb \pr \times \uu \pr} = \abs{\frac{E}{\Pm} \del_{H}^{2} \bb} \, .
    \label{eq:magnetic_balance_at_saturation}
\end{equation}

Using local flow length scale $\ell_{1}$ and magnetic field length scale $\ell_{2}$, \Equation{\ref{eq:magnetic_balance_at_saturation}} scales as

\begin{equation}
    \frac{\abs{\bld{B}} \abs{\uu}}{\ell_{1}} = \frac{E}{\Pm} \frac{\abs{\bb}}{\ell_{2}^{2}}
    \quad \Rightarrow \quad
    \frac{\abs{\bb}}{\abs{\uu}} = \frac{\abs{\bld{B}}}{\ell_{1}} \frac{\Pm \ell_{2}^{2}}{E} \, .
    \label{eq:magnetic_scaling}
\end{equation}

The \alfven\ timescale can be defined as $\tau_{A} = \ell_{1} / \ualf$, and the magnetic diffusion timescale as $\tau_{D} = \Pm \ell_{2}^{2} / E$.  The ratio $\mathrm{Lu} = \tau_{D} / \tau_{A}$ is then the Lundquist number.  Thus, by using $\abs{\uu} \approx \urms$, $\abs{\bb} \approx \brms$, and $\ualf = \abs{\bld{B}} \approx \backbs$, \Equation{\ref{eq:magnetic_scaling}} becomes

\begin{equation}
    \frac{\brms}{\urms} = \frac{\backbs}{\ell_{1}} \frac{\Pm \ell_{2}^{2}}{E} = \frac{\tau_{D}}{\tau_{A}} = \mathrm{Lu} \, .
    \label{eq:lundquist}
\end{equation}

Hence, the requirement of $\brms \gg \urms$ implies that we must be in a regime where $\mathrm{Lu} \gg 1$, which in turn places constraints on the combinations of $E$, $\Pm$ and $B_{0s}$ which we may choose.  For an Earth-like setup, we would ideally have $\Pm \ll 1$, which requires that we pick a sufficiently large ratio of $\backbs / E$.  In practice, however, numerical constraints limit both the maximum possible value of \backbs\ and the minimum possible value of $E$. In addition, it is desirable to keep $\ualf < 1$, so that the \alfven\ wave speed is smaller than the inertial wave speed and our model remains consistent with our QG assumption on rigid flows.  Thus, for a numerically achievable ratio of $\backbs / E$, while we may choose  $\Pm < 1$,  \Pm\ must remain sufficiently large such that solutions are in a regime characterized by $\mathrm{Lu} \gg1$.\\

The final aspect of the convective regime to be addressed concerns the influence of the Lorentz force $F_{\mathrm{L}}$ on the vorticity equation of \Equation{\ref{eq:vorz_nonaxi}}.  Left to evolve dynamically, flow and magnetic field lines tend to align themselves so as to limit induction by shear \citep[e.g.][]{Schaeffer+2017}. However, because we impose a steady background magnetic field in our model, such a fully dynamic reorganization is not possible.  As a result, for the parameters that we use (see next section), convective eddies are unduly constrained by the Lorentz force and become thin, elongated columns in the $s$-direction, sharing little resemblance with the large-scale flows in Earth's core. However, when we set $F_{L}=0$ in \Equation{\ref{eq:vorz_nonaxi}} and ignore the influence of the Lorentz force on the non-axisymmetric flows, we retrieve Earth-like large-scale eddies.   Because our main goal is to illustrate the mechanism by which Earth-like, large-scale flows can interact with the background field to generate slow zonal accelerations, our model is a better analog to Earth's core with $F_{\mathrm{L}}$ turned off.  In other words, although convective eddies in our model are still allowed to distort the magnetic field, we do not take into account the feedback of the Lorentz force on the flow eddies.  We retain the Lorentz torque \gammaL\ in the axially symmetric torque balance of \Equation{\ref{eq:vorz_axi}}, as it is our primary objective to show that it can be the dominant driver of rigid accelerations.


\section{Results}
\label{sec:results}
\subsection{Solution Scheme}
\label{sec:results:scheme}

A semi-spectral method with 768 Fourier modes in azimuth and radial derivatives approximated by second-order finite differences between 901 points arranged on a Chebyshev grid in radius was used to discretize \Equations{\ref{eq:vorz_nonaxi}}, (\ref{eq:vorz_axi}), (\ref{eq:induction_nonaxi}), (\ref{eq:induction_axi}), and (\ref{eq:heat}).  The resulting discrete equations were then evolved in timesteps of $5 \cdot 10^{-4}$ using a combination of a Crank-Nicolson method for the linear terms and a second-order Adams-Bashforth scheme for the non-linear terms \citep[e.g.][]{He+2007}.  Use of a Chebyshev grid ensured fine enough spacing to resolve boundary layers (proportional in thickness to $E^{1/2} \approx 0.002$), while maintaining reasonable computation times via coarser spacing away from the boundaries.  In our setup, minimum spacing near the boundaries is $2 \cdot 10^{-6}$, while maximum spacing in the centre of the model domain is $1 \cdot 10^{-3}$.  We set $s_{1} = 0.35$, mimicking the thickness ratio of Earth's core, and $s_{2} = 0.98$.  Limiting our domain to $s_{2} =0.98$ instead of $s_{2} =1$ is convenient, as it allows us to use a slightly larger grid space and timesteps (since $\beta \rightarrow \infty$ as $s \rightarrow 1$).  The results that we now present come from an experiment with $E = 5.0 \cdot 10^{-6}$, $R_{a} = 5.0 \cdot 10^{8}$, $\backbs = 0.15$, $P_{m} = 0.1$, and $P_{r} = 1.0$.  With these parameters, a numerical simulation started from small perturbations typically takes about 1000 non-dimensional time units (or about 160 full rotation) to reach statistical equilibrium.  

\subsection{General features of the flow and induced magnetic fields}
\label{sec:results:general_features}

\begin{figure}
    \centering 
    \begin{minipage}{0.475\textwidth}
            \includegraphics[width=\textwidth, keepaspectratio]
                            {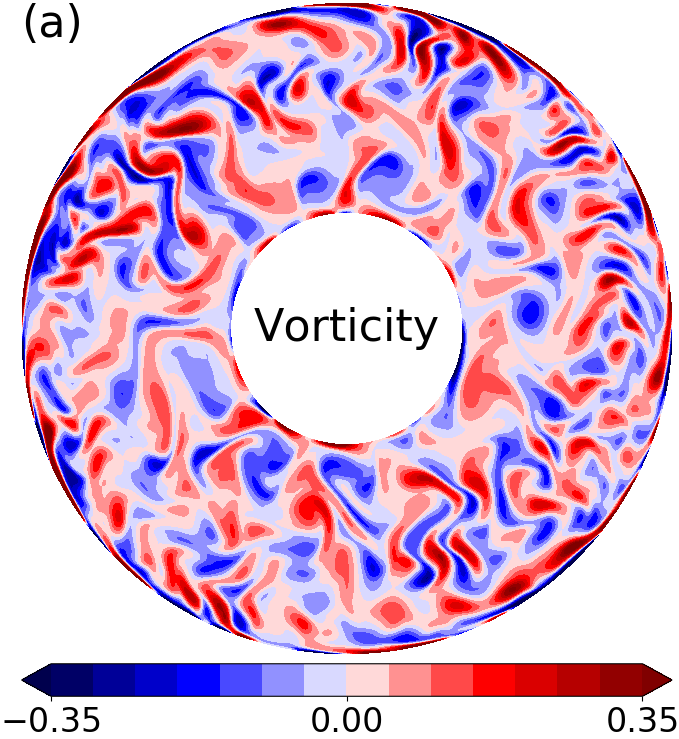}
        \end{minipage} 
        \begin{minipage}{0.475\textwidth}
            \includegraphics[width=\textwidth, keepaspectratio]
                            {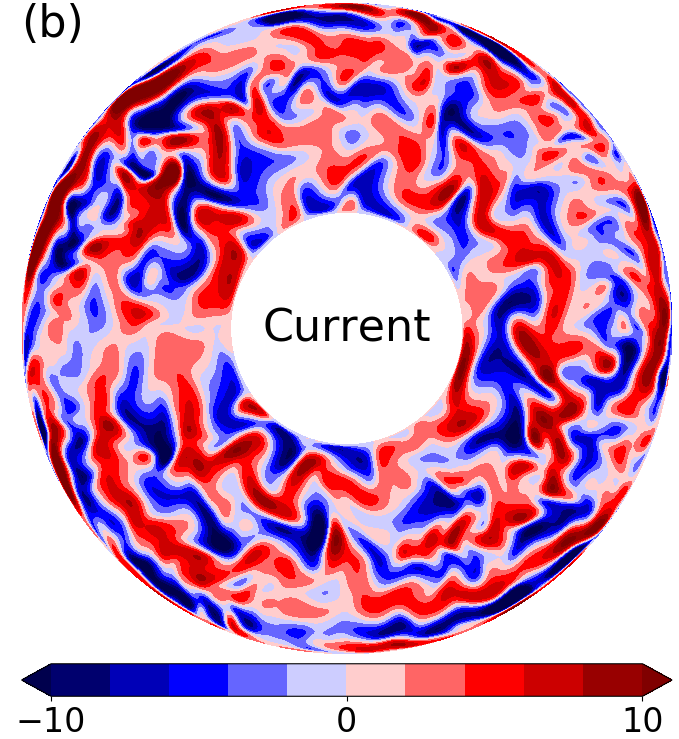}
        \end{minipage}
        \hfill
        \caption{\small Snapshots in time of (a) nonaxisymmetric vorticity $\pwz$ and (b)
      nonaxisymmetric axial current $\pjz$, as seen looking downward from the north pole.}
    \label{fig:overhead_52}
\end{figure}

Fig.~\ref{fig:overhead_52} shows a snapshot in time of the nonaxisymmetric \pwz\ and \pjz\ after the
model's global energy budget has reached statistical equilibrium.  The presence of eddies in both
the flow and the perturbed magnetic field are clearly visible.  The magnetic field perturbations
exhibit structures with larger wavelengths than those in the flow field, and with smoother features,
a result of the more rapid magnetic to viscous diffusion ($P_{m} = 0.1$).\\

Magnetic field perturbations result from the action of convective eddies shearing and advecting the
sum of the background and perturbed magnetic field.  Time-dependency in both the flow and the
magnetic field leads to fluctuations in the Reynolds ($\gammaR$, Eq.~\ref{eq:reynolds_axi}) and
Lorentz ($\gammaL$, Eq.~\ref{eq:lorentz_axi_main}) torques, respectively.  These must be
accommodated by zonal accelerations.\\

Typical \urms\ and \brms\ values after equilibration are $0.0024$ and $0.14$, respectively.  The
amplitude of magnetic field perturbations in our numerical experiment is of the same order as the
imposed background field, $\abs{\bb} \approx \abs{\bld{B}_{0}}$.  The \alfven\ and Lundquist numbers
of our simulation are then $\alfnum \approx 0.014$ and $\mathrm{Lu} \approx 70$, within the region
of parameter space where we expect $\gammaL \gg \gammaR$.  This is the region where the decadal
timescale dynamics of zonal accelerations in Earth's core should reside.

\subsection{Time-averaged axisymmetric force balance}
\label{sec:results:axisymmetric_forces}

\begin{figure}
    \centering
             {\includegraphics[width=0.9\textwidth,
                               keepaspectratio]
                              {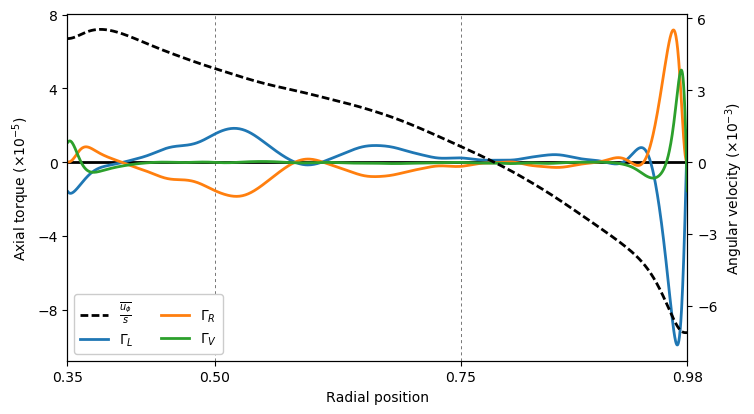}}
    \caption{Time-averaged mean axial torques (solid lines) and time-averaged zonal angular
      velocity (dashed line) as a function of radius.}
    \label{fig:time_averages}
\end{figure}

The dashed black line of Fig.~\ref{fig:time_averages} shows the time-averaged zonal angular
velocity.  Its profile is dominated by a shear flow spanning the whole of the modeled region,
retrograde at the outer boundary and prograde at the inner boundary. The amplitude of this shear
flow is of the same order of magnitude as the amplitude of typical convective eddies. \\

The mean (time-averaged) zonal flow of Fig.~\ref{fig:time_averages} differs from the one typically observed in the
absence of a magnetic field.  In non-magnetic convection with stress-free boundaries, the direction
of the mean zonal flow is reversed -- it is prograde at the equator -- and its characteristic velocity is
much larger than the velocities associated with convective eddies.  It results from time-averaged
Reynolds stresses, themselves being the product of the topographic beta effect acting on convective
eddies \citep[e.g.][]{Cardin1994,christensen2002}. \\

In the presence of a radial magnetic field, distortion of this magnetic field by the shear flow induces a restoring Lorentz force (through \gammaLone\ of Eq.~\ref{eq:lorentz_axi_1}) which limits the growth of the mean zonal flow.  In three-dimensional models, this mean zonal flow is not $z$-invariant \citep[e.g.][]{aubert2005}.  In our QG model, because our magnetic field perturbation is defined with a built-in topographic beta effect identical to that of the flow, a time-averaged Maxwell stress (\gammaLtwo, Eq.~\ref{eq:lorentz_axi_2}) is maintained in the same way as the time-averaged Reynolds stress \gammaR.  Since \gammaLtwo\ has the same form, but opposite sign, as \gammaR, and since the magnitude of the former typically dominates that of the latter in our model, the direction of
the driven mean zonal flow is reversed to that produced in non-magnetic convection.\\

\begin{figure}
    \centering
                 {\includegraphics[width=0.9\textwidth,
                               keepaspectratio]
                              {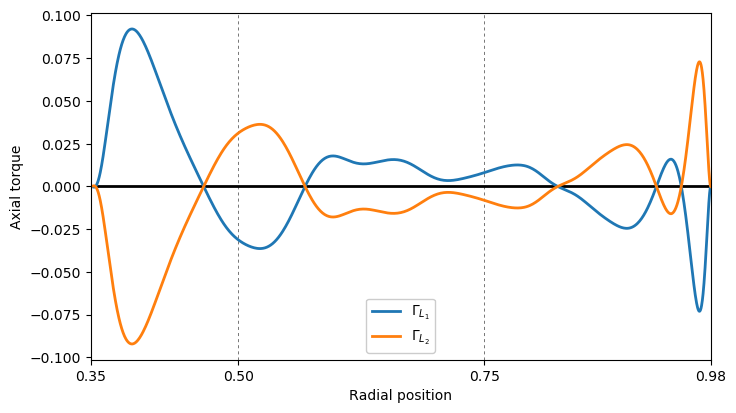}}
             \caption{Time-averaged components of the mean axial Lorentz torque as a function of radius.}
    \label{fig:magnetic_force_balance}
\end{figure}

The time-averaged $\overline{u_\phi}$ profile, then, is the result of a balance between the time-averaged torques.  Fig.~\ref{fig:time_averages} illustrates this balance, showing the time-averaged Reynolds (\gammaR, orange), Lorentz (\gammaL, blue), and viscous (\gammaV, green) torques as a function of radius.  In the interior of the domain, the Reynolds and Lorentz torques largely balance one another. The viscous torque becomes more important near the boundaries, especially the outer one.  The enhancement of all three torques near the outer boundary is caused by the large $\beta$-effect: both the viscous and the \gammaLone\ part of the Lorentz torque depends explicitly on $\beta$ (through the $s$-derivative of $L$, see Eqs. \ref{eq:viscous_axi} and \ref{eq:lorentz_axi_1}), while the Reynolds (\gammaR, \Equation{\ref{eq:reynolds_axi}}) and Maxwell (\gammaLtwo, \Equation{\ref{eq:lorentz_axi_2}}) torques implicitly involve $\beta$ through their dependence on \pup\ and \pbp, respectively.  We further note that both \gammaLone\ and \gammaLtwo\ are individually much larger than \gammaL, as shown by Fig. \ref{fig:magnetic_force_balance}.  However, they tend to cancel one another, leaving a net Lorentz torque several orders of magnitude smaller than either one individually.  This self-cancellation will be re-examined in Section \ref{sec:results:taylorization}.

\subsection{Zonal accelerations}
\label{sec:results:zonal}

\begin{figure}
    \centering
             {\includegraphics[width=0.900\textwidth, keepaspectratio]{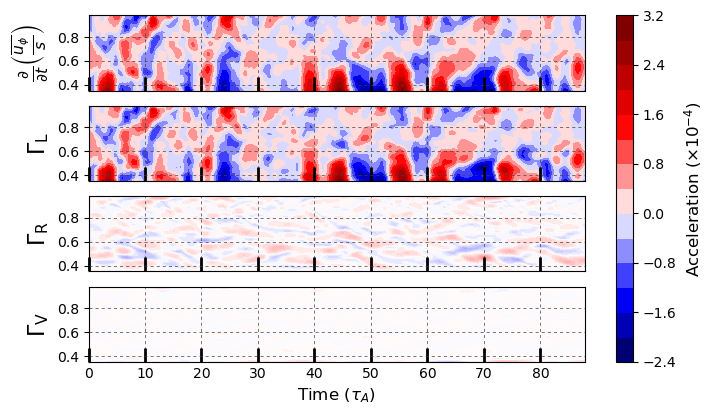}}
    \caption{The angular acceleration (top panel), and the time-dependent parts of the
      Lorentz (second panel), Reynolds (third panel) and viscous (bottom panel) torques, as
      functions of cylinder radius and time.  The time-averaged contribution to each torque (shown
      in Fig. \ref{fig:time_averages}) has been removed.}
    \label{fig:forceplot}
\end{figure}

Fluctuations in time with respect to this time-averaged torque balance are the main focus of our study.
Fig.~\ref{fig:forceplot} shows the axisymmetric angular accelerations (top panel) and the
time-varying parts of \gammaL, \gammaR, and \gammaV\ (bottom three panels) after the system has
equilibrated.  Fluctuations in \gammaL\ have a typical amplitude five times larger than those of
\gammaR: $\sim 6.6\times10^{-5}$ versus $\sim 1.2\times10^{-5}$.  \gammaV\ only plays a small role in the time-dependent dynamics, with RMS fluctuations of the order of $\sim
0.3\times10^{-5}$.\\

Thus, fluctuations in our model's zonal accelerations are mainly controlled by the Lorentz torque.
Indeed, the upper two panels of Fig.~\ref{fig:forceplot} suggest a strong correlation at all times
and radii between $\pdt\pl\vorO\pr$ and \gammaL.  Time on Fig.~\ref{fig:forceplot} has been scaled
to the timescale of the fundamental \alfven\ wave mode, $\tau_{A} = 2 (s_{2} - s_{1}) /
\ualf$.  The joint fluctuations in the zonal accelerations and Lorentz torque cover a broad range of timescales but are dominated by periods which fall in the range of $5$ to $10$ times longer than $\tau_{A}$.   This slower timescale, $\tau_{\mathrm{slow}} \approx 5-10\, \tau_{A}$, reflects the time-fluctuations of the magnetic field, themselves
the result of induction by convective flows. These occur on a longer timescale than the \alfven\ wave
propagation timescale, as we expect for a regime with $\alfnum < 1$.  Scaling the temporal fluctuations on Fig.~\ref{fig:forceplot} to Earth's
core, taking $\tau_{A} \approx$ 6 yr, gives a timescale $\tau_{\mathrm{slow}}$ of 30 to 60 yr for these
magnetically-driven zonal accelerations, similar to the zonal accelerations
inferred within Earth's core.\\


To further demonstrate that the slow magnetic field fluctuations originate from the underlying
convective dynamics, we compute a characteristic azimuthal wave number $k^{*}$ \citep[e.g.][]{takahashi2008} at each radius $s$ from the convolution of wavenumber $k$ and convective speed $u(k)$:

\begin{equation}
    k^{*}(s) = \frac{\integral{}{}{k u(k)}{k}}{\integral{}{}{u(k)}{k}} \, .
\end{equation}

\begin{figure}
    \centering
    \includegraphics[width=1.0\textwidth,
                     keepaspectratio]
                    {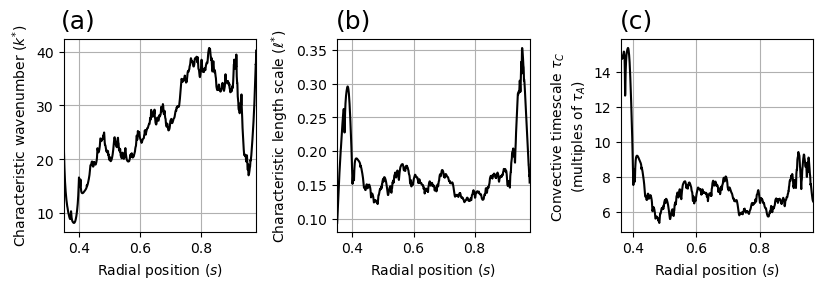}
    \caption{Characteristic (a) wave numbers, (b)
      length scales, and (c) time scales as a function of radius at a given snapshot in time.}
    \label{fig:characteristic_scales}
\end{figure}

The result of this calculation as a function of radius, for the snapshot corresponding to time equal 0 on Fig.~\ref{fig:forceplot}, is shown in panel (a) of Fig.~\ref{fig:characteristic_scales}.  The characteristic convective length scale $\ell_{C}(s)$ is then
calculated from the characteristic wavenumber as

\begin{equation}
    \ell_{C}(s) = \frac{2 \pi s}{k^{*}(s)} \, .
\end{equation}

The characteristic convective length scale for all radial positions is shown in panel (b) of Fig.~\ref{fig:characteristic_scales}.  Dividing these length scales by the RMS velocities at each radius
yields a characteristic convective timescales $\tau_{C}$, which is representative of the time required for the
flow to create significant changes in the magnetic field:

\begin{equation}
    \tau_{C}(s) = \frac{\ell_{C}(s)}{\urms}  \, .
\end{equation}

The dependence of $\tau_{C}$ with radius is shown in panel (c) of Fig.~\ref{fig:characteristic_scales}.  All three plots of Fig.~\ref{fig:characteristic_scales} change for different time snapshots, but are representative of the general behaviour at all times. Furthermore, they only show the dominant length and time scales of the flow, which is truly characterized by a spectrum of scales.  As shown in Fig.~\ref{fig:characteristic_scales}c, in the bulk of the fluid $\tau_{C}$ is approximately equal to $5-8 \, \tau_{A}$, within the same range as $\tau_{\mathrm{slow}}$.  These are obviously simple order of magnitude calculations that do not capture the complex non-linear connection between large-scale flows and time changes in the Lorentz torque.  Nevertheless, the combination of Figs.~\ref{fig:forceplot} and \ref{fig:characteristic_scales} show that it is possible to drive slow zonal accelerations by fluctuating Lorentz torques, themselves driven by underlying convective flows shearing and advecting the magnetic field.\\

\begin{figure}
    \centering
             {\includegraphics[width=0.900\textwidth, keepaspectratio]{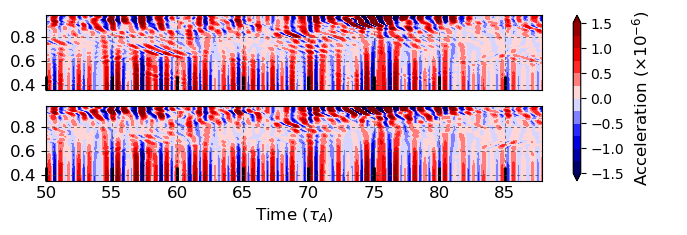}}
                 \caption{Output of a highpass filter applied to the second half of the zonal angular
      accelerations (top panel) and Lorentz torques (bottom panel) of Fig.~\ref{fig:forceplot}.
      In our non-dimensional time units, the 8th-order digital Butterworth filter's -3 dB frequency
      is 1/10.}
          \label{fig:filtered}
    \end{figure}

In addition to the slow fluctuations of Fig.~\ref{fig:forceplot}, the convective dynamics also
generate free \alfven\ waves.  Fig.~\ref{fig:filtered} shows the second half of the zonal
acceleration and Lorentz torque panels of Fig.~\ref{fig:forceplot} after applying a highpass
filter to remove fluctuations with periods longer than 1.19 $\tau_{A}$.  This reveals the presence
of periodic fluctuations with a period of approximately 1 $\tau_{A}$.  The correlation between the
acceleration and Lorentz torque shows that these are indeed \alfven\ waves.  (Applying the same
highpass filter to \gammaR\ and \gammaV\ produces only low-amplitude noise.)\\

Left on their own, free \alfven\ oscillations should decay away because of ohmic dissipation.  In
our model, they are continuously re-excited by the underlying convective dynamics, though resonant
amplification remains modest and their amplitude does not rise much higher than that resulting from
the forced background accelerations.  Their typical RMS velocities of $1 \cdot 10^{-4}$ are
approximately 7 times smaller than the RMS velocities of $7 \cdot 10^{-4}$ found for the slower
zonal flow fluctuations.  This is the reason why \alfven\ waves, though present, are not apparent on
Fig.~\ref{fig:forceplot}.\\

\subsection{Taylorization}
\label{sec:results:taylorization}

For a system in a perfect Taylor state, even though the Lorentz torque at any point may be large,
cancellations occur such that the Lorentz torque integrated over a geostrophic cylinder is equal to
zero.  The ``Taylorization'' is a measure of the degree of this Lorentz torque cancellation.  In our model, where the magnetic field (and therefore the Lorentz torque) is
assumed to be axially invariant, the Taylorization is measured by the factor $\mathcal{T}$:

\begin{equation}
    \mathcal{T}(s)
        = \frac{\abs{\cintegral{}{}{\mathcal{M}_{\phi}(s,\phi)}{\phi}}}{\cintegral{}{}{\abs{\mathcal{M}_{\phi}(s,\phi)}}{\phi}}
        = \frac{ \abs{\gammaL(s)} }{ \cintegral{}{}{\abs{\mathcal{M}_{\phi}(s,\phi)}}{\phi} } \, ,
    \label{eq:taylorization}
\end{equation}
where $\mathcal{M}_{\phi}(s,\phi)$ represents the azimuthal component of the local magnetic force.
A system with low Taylorization has $\mathcal{T} \lesssim 1$, while a system with a high
Taylorization has $\mathcal{T} \ll 1$.  Typical geodynamo simulations achieve $\mathcal{T} \approx
10^{-1} - 10^{-3}$ \citep[e.g.][]{rotvig2002,wicht2010,teed2014}.\\

\begin{figure}
\centering
{\includegraphics[width=0.900\textwidth, keepaspectratio]{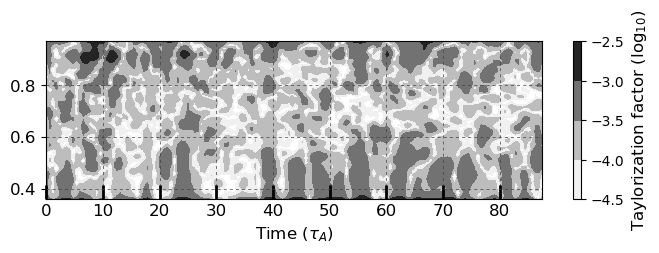}}
\caption{Taylorization factor $\mathcal{T}$ as a function of radius and time.  With our choice of boundary conditions, the Taylorization factor equals 1 at the boundaries.  Only $s \in [0.36,0.97]$ is shown.}
\label{fig:taylorization}
\end{figure}

Fig. \ref{fig:taylorization} shows the Taylorization factor of our model as a function of
cylindrical radius and time.  Typical RMS values of $\mathcal{T}$ are approximately $3.2 \cdot
10^{-4}$, with peak values of $2.6 \cdot 10^{-3}$.  As shown in the previous section, temporal
fluctuations of the magnetic field lead to temporal fluctuations of the Lorentz torque.  Thus,
Taylor's constraint is continually broken, with the Lorentz torque fluctuations being accommodated
by rigid zonal accelerations.  This can be observed in Fig. \ref{fig:forceplot}, where the times
and radii of the largest zonal accelerations often coincide with the largest values of $\mathcal{T}$
in Fig. \ref{fig:taylorization}.\\

Therefore, our model is not in a perfect Taylor state at any given moment, but instead fluctuates
about an equilibrium state characterized by a low Taylorization factor.  A part of the cancellation of the
Lorentz torque over any given geostrophic cylinder is inherent to the form of the Maxwell torque
(\gammaLtwo), as it involves products of the magnetic field vectors which change direction at
different azimuthal points. On a time-average, \gammaLtwo\ does not vanish but it is mostly opposed by  \gammaLone, as shown by Fig. \ref{fig:magnetic_force_balance}.  Yet, \gammaLone\ and \gammaLtwo\ do not cancel exactly: the time-averaged Taylorization factor is small ($\mathcal{T} \sim 10^{-4}$) but not zero.  In our model, the departure from a time-averaged Taylor state is dominantly caused by the persistent torque from Reynolds stresses (and from viscous forces near the domain boundaries, see Fig. \ref{fig:time_averages}).\\

\section{Discussion}
\label{sec:discussion}
\subsection{Slow Zonal Accelerations and Decadal Timescale Dynamics in Earth's Core }
\label{sec:discussion:slow}

Our model shows that in an Earth-like parameter regime characterized by $\alfnum \ll 1$ and
$\mathrm{Lu} \gg 1$, convective flows can drive fluctuations in the magnetic field which then lead
to temporal fluctuations in the Lorentz torque.  The latter generate zonal accelerations in the form
of free \alfven\ waves, but also slower forced zonal accelerations on a timescale connected to the
convective flows.  In order to drive large-scale, core-size, zonal accelerations, convective eddies must be large-scale themselves.  Because the typical velocity of these large-scale eddies is slower than the \alfven\ wave velocity, the timescale associated with the magnetic field change they produce, and thus the Lorentz torque, is slower than $\tau_{A}$, the timescale of \alfven\ wave propagation across the core.\\

In our model, the typical convective timescale associated with large-scale eddies is 5 to 8 times longer than $\tau_{A}$.  These eddies induce Lorentz torque fluctuations over a broad range of timescales, but dominated by periods that are 5 to 10 times longer than $\tau_{A}$.  Scaling our results to Earth's core by taking $\tau_{A} = 6$ yr, our model produces zonal accelerations with typical timescales in the range of 30 to 60 yr, comparable to observations.  Furthermore, the power spectrum of flows reconstructed from geomagnetic secular variation shows peaks at spherical harmonic degrees 8-10 \cite[e.g. Fig. 5 of][]{gillet2015}. At mid-core radius, this corresponds to a typical length scale of approximately 600 km.  Taking a typical velocity of 10 km yr$^{-1}$, these are associated with a typical convective timescale of approximately 60 yr.  If these flows rigidly extend axially through the core, according to our mechanism they should distort the magnetic field, and thus produce fluctuations in the Lorentz torque and zonal accelerations in response, at approximately the same 60 yr timescale.   Our results suggest that the inferred decadal, large-scale zonal accelerations of geostrophic cylinders in the Earth's core can be explained by forced fluctuations of the Lorentz torque, themselves driven by the large-scale convective flows.\\

As observed in Earth's core, our dynamical model contains large-scale flows, slow zonal accelerations, and faster \alfven\ waves.  In our model, their typical non-dimensional velocities are, respectively, $2.4 \cdot 10^{-3}$, $7 \cdot 10^{-4}$ and $1\cdot 10^{-4}$, for a relative ratio between them of 34:7:1.  In Earth's core, these typical velocities are 10 km yr$^{-1}$, 2 km yr$^{-1}$ and 0.2 km yr$^{-1}$ \cite[e.g.][]{gillet2015}, for a ratio of 50:10:1.   Therefore, although their amplitude ratios do not match those of Earth's core exactly, our model reproduces the correct ordering between these flows.  This gives us further confidence that our model is a good analog for the decadal flow dynamics in Earth's core.\\

Although our dynamical model is in the correct parameter regime in terms of $\alfnum \ll 1$ and
$\mathrm{Lu} \gg 1$, other parameters remain far from Earth-like, notably the Ekman number and
magnetic Prandtl number.  Care must then be taken when extrapolating our results to Earth's core.
To further confirm that our results appropriately capture the decadal timescale dynamics of zonal
flows in the core, it would be desirable to carry many more numerical experiments, systematically
varying some of the input parameters in order to develop scaling properties for our model.  We plan
to do this in a future study.  A different approach is to try to do a similar analysis in a 3D
geodynamo model, some of which are approaching the parameter regime of $\alfnum \ll 1$ and
$\mathrm{Lu} \gg 1$ which we have highlighted \citep[e.g.][]{Aubert2017, Schaeffer+2017}.  Short of
doing this, we may speculate on how the use of more Earth-like Ekman and magnetic Prandtl numbers
would affect the dynamics of zonal flows.\\

For the relatively large Ekman number of our numerical experiment, viscous forces remain important in the establishment of the large-scale, quasi-core-size eddies that are visible on
Fig.~\ref{fig:overhead_52} \cite[e.g][]{aurnou2015}.  However, at  $E\approx10^{-15}$, as inferred for
Earth's core, the $E^{1/3}$ scaling law suggests that the typical length scale of
viscously-controlled eddies should be much smaller, about $10^{-5}$ times the core size, or of the
order of 10-100 m in width.  Such small eddies would be inefficient at generating the core-size
changes in the magnetic field that are required to drive large-scale zonal accelerations.\\

However, large-scale eddies are present in Earth's core, as inferred from the geomagnetic secular variation.  The typical length-scale of eddies in the core must then be controlled by processes other than this viscous scaling.  Two main reasons can be invoked. First, for sufficiently low Ekman numbers, small-scale structures in the core likely feed their energy to larger length scales via an inverse energy cascade driven by geostrophic turbulence \cite[e.g.][]{guervilly2014,stellmach2014}. Second, the influence of the Lorentz force should also lead to larger convective length scales \cite[e.g.][]{roberts13}, a regime which is beginning to be accessible in numerical simulations  \citep[e.g.][]{Matsui+2014, Yadav+2016, Schaeffer+2017}.    A combination of these two effects may be responsible for forming the large structures seen in the core.  Both are absent in our model.\\

Therefore, although the dynamics that sustain the large-scale eddies in our numerical model are most likely not Earth-like, the large-scale eddies themselves share some Earth-like qualities.  The key ingredient for driving decadal zonal accelerations according to our mechanism is the very presence of large-scale flow eddies with decadal convective timescale, not precisely how these eddies are generated.\\

The magnetic Prandtl number in the core is much lower than the one we have chosen in our numerical
experiment.  Since a lower \Pm\ corresponds to enhanced magnetic diffusion, temporal changes in the
magnetic field would be dominantly controlled by the largest length scales of the underlying
convective flow.  Thus, while small-scale eddies certainly exist in the core, changes in the
magnetic field should still preferentially occur at the largest length scale.  We therefore expect that
temporal changes in the Lorentz torque can drive decadal zonal accelerations at core-size wavelengths
in the radial direction, just as we observe in our numerical experiment of limited spatial
resolution.

\subsection{Free \alfven\ waves}
\label{sec:discussion:fast}

As shown in Fig.~\ref{fig:filtered}, our dynamical model excites free \alfven\ waves.  However, their spatio-temporal properties  differ from those detected in Earth's core.  In our
model, they are dominated by a standing wave oscillation of the fundamental mode.  In Earth's core,
they take the form of outward travelling waves \cite[][]{gillet2010,gillet2015}.\\

The reason why \alfven\ waves in Earth's core travel outward, as well as their excitation mechanism,
remain unclear.  Outward travelling \alfven\ waves resulting from a quasi-periodic triggering near
the tangent cylinder are observed in some numerical models \cite[][]{teed2014, teed2015,
  Schaeffer+2017}.  When approaching an Earth-like regime, Lorentz forces are responsible for this
torque, but the precise physical mechanism has not been clearly identified.\\

It would be a valuable effort to investigate where and how \alfven\ waves are excited in our model.
Since we do not see a preferential propagation direction, excitation appears to be distributed
evenly within the integration domain of our model. The region close to the tangent cylinder does not
appear to be the seat of of any form of recurrent instability, although this may be because we have
not modelled the dynamics inside the tangent cylinder.\\

Given that convective flows in our model induce changes in the magnetic field on a broad range of
timescales, \alfven\ waves on Fig.~\ref{fig:filtered} may simply represent the resonant response to
fluctuations of the Lorentz torque which occur in the vicinity of their free period range.  Indeed,
correlations between Figs.~\ref{fig:filtered} and \ref{fig:taylorization} suggest that this is the
case: notable increases in the Taylorization factor are often, though not always, associated with an
amplitude enhancement of free \alfven\ waves.  This argues along the same line of a recent study
which has shown that applying a stochastic forcing in the volume of the core readily excites
\alfven\ waves \cite[][]{gillet2017}.  Moreover, the study of \citet{gillet2017} has also shown that
electromagnetic dissipation at the CMB transforms standing \alfven\ waves into outward traveling
waves, with similar characteristics as those detected in the Earth's core.  Our model constitutes a
dynamical realization of such a stochastic forcing and supports the idea that \alfven\ waves in
Earth core are simply the response to sub-decadal changes in the Lorentz torque within the bulk of
the core.  Our model does not include dissipation at the CMB, but we believe (though this should be
tested) that adding it would also transform our standing \alfven\ waves into outwardly propagating
waves.

\subsection{Taylorization}
\label{sec:discussion:taylorization}

Taylor's constraint is continuously being broken in our model, generating zonal accelerations in
response.  The numerical value of the Taylorization factor associated with these fluctuations is of
the order of $10^{-3}$, similar to that extrapolated to Earth's condition in recent geodynamo models
\citep{Aubert2017}.  The time-average state (or statistical equilibrium) about which these fluctuations occur is characterized by a higher degree of Taylorization, with a Taylorization factor of the order of $10^{-4}$.  Though this is small, it indicates that the equilibrium state in our model remains far from a perfect Taylor state. In the bulk interior, this is dominantly because Reynolds stresses lead to a non-zero time-averaged torque.  Thus, the time-averaged Lorentz torque need not be zero (and obey a Taylor state), but only be as small as the persistent torque from Reynolds stresses.\\

The results of our model suggest that, as was pointed out by \citet{dumberry2003}, the torque from
Reynolds stresses in Earth's core may also play a leading order role in the departure from a
time-averaged Taylor state. The ratio of the torque from Reynolds stresses to the Lorentz torque scales as  $(u_C/u_A)^2$, so it is proportional to the square of the  \alfven\ number.  Because the \alfven\ number in our model is similar to that in Earth's  core, the baseline Taylorization factor of approximately $10^{-4}$ in Fig.~\ref{fig:taylorization} may also be representative of that expected in Earth's core.  However, because our model is two-dimensional, extrapolating our results to the three-dimensional magnetic field of Earth is clearly not straightforward.  Furthermore, Ekman friction, which we have neglected, could be important in the time-averaged torque balance, especially if turbulent processes lead to an enhanced effective viscosity.  Ekman friction could balance a part of the torque from Reynolds stresses and the Taylorization factor could then be smaller than $10^{-4}$.   Regardless of its exact value, the high degree of Taylorization requires that large cancellations in the equilibrium Lorentz torque over a cylinder must occur in Earth's core.  It this sense, exploring dynamo solutions in the limit of a vanishing Lorentz torque remains a worthy goal \cite[e.g.][]{livermore2008,wu2015}.  This being said, one must keep in mind that the correct Taylorization factor in Earth's core may not be asymptotically close to zero but instead be closer to $10^{-4}$.

\section{Conclusion}
\label{sec:conclusion}

We have constructed a two-dimensional reduced model of rotationally-dominated magnetoconvection
capable of producing distinct short- and long-timescale accelerations in the zonal flow.  The
short-timescale accelerations are free \alfven\ waves, while the long-timescale accelerations are
magnetically forced through the evolution of the Lorentz torque.  The temporal changes in the
magnetic field which drive the time-varying Lorentz torque are produced by the underlying convective
flows, shearing and advecting the magnetic field on a timescale associated with convective eddies.
Our results provide a dynamical explanation for the rigid decadal zonal accelerations that are
inferred to exist in Earth's core on the basis of the magnetic field observations and changes in
LOD.\\

Our results offer an alternative to the recent suggestion that the decadal zonal accelerations may
not reflect deep seated rigid flows but are instead free Magnetic-Archemedian-Coriolis (MAC) waves
in a stratified layer at the top of the core \cite[][]{buffett2014,buffett2016,jaupart17}.  The zonal flows of
such MAC waves are characterized by a shear in the radial direction: flow at the CMB does not
reflect flow deeper in the core in the axial direction. It is then more difficult to build a
prediction of LOD changes based solely on the knowledge of flows at the CMB.  It is nevertheless possible that, when properly taking into account the coupling of flows in the bulk of the core with
these MAC waves, a prediction of core angular momentum change may be constructed so as to match the
observed LOD variations \cite[][]{buffett2016}.  Yet the very fact that a very good match exists between
the observed changes in the LOD and those predicted on the basis of purely rigid zonal flows
suggests that deviations from rigidity are limited.  Moreover, as our dynamical model shows,
convective dynamics are expected to drive temporal changes in deep seated rigid zonal flows at
decadal timescales.  If the top of the core is stably stratified, rigid zonal flows may drive forced
MAC oscillations and/or excite free MAC waves.  Hence, the zonal flows at the CMB may consist of a
combination of deep seated rigid zonal flows and flows that obey a MAC balance.  However, because of the good match in LOD based on purely rigid flows, we speculate that the non-rigid MAC flows, if
present, make up only a fraction of the total zonal flow.

\begin{acknowledgments}
This work was supported by both a NSERC/CRSNG Post-Graduate Scholarship and a Discovery grant from NSERC/CRSNG.  Numerical simulations were performed on computing facilities provided by WestGrid and Compute/Calcul Canada.  Figures were created using the Matplotlib package for          Python \citep{Hunter2007}.  We thank Nathana\"el Schaeffer and an anonymous reviewer for their constructive comments and suggestions.  We further express our gratitude to Nathana\"el Schaeffer, who kindly shared a numerical QG code which we extended over the course of this project.
\end{acknowledgments}

\bibliographystyle{gji}
\bibliography{more_dumberry_gji18_arxiv}

\begin{thebibliography}{56}
\expandafter\ifx\csname natexlab\endcsname\relax\def\natexlab#1{#1}\fi

\bibitem[Aubert(2005)]{aubert2005}
Aubert, J., 2005.
\newblock Steady zonal flows in spherical shell dynamos, {\it J. Fluid
  Mech.\/}, {\bf 542}, 53--67.

\bibitem[Aubert et~al.(2003)Aubert, Gillet, \& Cardin]{Aubert2003}
Aubert, J., Gillet, N., \& Cardin, P., 2003.
\newblock Quasigeostrophic models of convection in rotating spherical shells,
  {\it Geochem. Geophys. Geosyst.\/}, {\bf 4}, 1052.

\bibitem[Aubert et~al.(2017)Aubert, Gastine, \& Fournier]{Aubert2017}
Aubert, J., Gastine, T., \& Fournier, A., 2017.
\newblock Spherical convective dynamos in the rapidly rotating asymptotic
  regime, {\it J. Fluid Mech.\/}, {\bf 813}, 558--593.

\bibitem[Aurnou et~al.(2015)Aurnou, Calkins, Cheng, King, Nieves, Soderlund, \&
  Stellmach]{aurnou2015}
Aurnou, J., Calkins, M.~A., Cheng, J.~S., King, E.~M., Nieves, D., Soderlund,
  K.~M., \& Stellmach, S., 2015.
\newblock Rotating convective turbulence in {E}arth and planetary cores, {\it
  Phys. Earth Planet. Inter.\/}, {\bf 246}, 52--71.

\bibitem[Braginsky(1970)]{Braginsky1970}
Braginsky, S.~I., 1970.
\newblock Torsional magnetohydrodynamic vibrations in the {E}arth's core and
  variations in day length, {\it Geomag. Aeron.\/}, {\bf 10}, 1--10.

\bibitem[Buffett(2010)]{buffett2010}
Buffett, B.~A., 2010.
\newblock Tidal dissipation and the strength of the earth's internal magnetic
  field, {\it Nature\/}, {\bf 468}, 952--954.

\bibitem[Buffett(2014)]{buffett2014}
Buffett, B.~A., 2014.
\newblock Geomagnetic fluctuations reveal stable stratification at the top of
  the {E}arth's core, {\it Nature\/}, {\bf 507}, 484--487.

\bibitem[Buffett et~al.(2016)Buffett, Knezek, \& Holme]{buffett2016}
Buffett, B.~A., Knezek, N., \& Holme, R., 2016.
\newblock Evidence for {MAC} waves at the top of earth's core and implications
  for variations in length of day, {\it Geophys. J. Int.\/}, {\bf 204},
  1789--1800.

\bibitem[Busse \& Or(1986)]{busse1986a}
Busse, F. \& Or, A., 1986.
\newblock Convection in a rotating cylindrical annulus: thermal rossby waves,
  {\it J. Fluid Mech.\/}, {\bf 166}, 173--187.

\bibitem[Cardin \& Olson(1994)]{Cardin1994}
Cardin, P. \& Olson, P., 1994.
\newblock Chaotic thermal convection in a rapidly rotating spherical shell:
  Consequences for flow in the outer core, {\it Phys. Earth Planet. Inter.\/},
  {\bf 82}, 235--259.

\bibitem[Chao et~al.(2014)Chao, Chung, Shih, \& Hsieh]{Chao2014}
Chao, B.~F., Chung, W.~Y., Shih, Z.~R., \& Hsieh, Y.~K., 2014.
\newblock Earth's rotation variations: a wavelet analysis, {\it Terra Nova\/},
  {\bf 26}, 260--264.

\bibitem[Christensen(2002)]{christensen2002}
Christensen, U., 2002.
\newblock Zonal flow driven by strongly supercritical convection in rotating
  spherical shells, {\it J. Fluid Mech.\/}, {\bf 470}, 115--133.

\bibitem[Christensen \& Wicht(2015)]{Christensen2015}
Christensen, U. \& Wicht, J., 2015.
\newblock Numerical dynamo simulations, in {\em Treatise on Geophysics\/},
  chap. 8.10, pp. 245--277, ed. Schubert, G., Elsevier, Oxford.

\bibitem[Christensen \& Aubert(2006)]{christensen2006a}
Christensen, U.~R. \& Aubert, J., 2006.
\newblock Scaling properties of convection-driven dynamos in rotating spherical
  shells and application to planetary magnetic fields, {\it Geophys. J.
  Int.\/}, {\bf 166}, 97--114.

\bibitem[Cowling(1957)]{Cowling1957}
Cowling, T., 1957.
\newblock The dynamo maintenance of steady magnetic fields, {\it Q. J. Mech.
  Appl. Math.\/}, {\bf 10}, 129--136.

\bibitem[Dumberry \& Bloxham(2003)]{dumberry2003}
Dumberry, M. \& Bloxham, J., 2003.
\newblock Torque balance, taylor's constraint and torsional oscillations in a
  numerical model of the geodynamo, {\it Phys. Earth Planet. Inter.\/}, {\bf
  140}, 29--51.

\bibitem[Finlay et~al.(2010)Finlay, Dumberry, Chulliat, \& Pais]{Finlay2010}
Finlay, C.~C., Dumberry, M., Chulliat, A., \& Pais, M., 2010.
\newblock Short timescale core dynamics: Theory and observations, {\it Space
  Sci. Rev.\/}, {\bf 155}, 177--218.

\bibitem[Gillet \& Jones(2006)]{Gillet2006}
Gillet, N. \& Jones, C.~A., 2006.
\newblock The quasi-geostrophic model for rapidly rotating spherical convection
  outside the tangent cylinder, {\it J. Fluid Mech.\/}, {\bf 554}, 343--370.

\bibitem[Gillet et~al.(2010)Gillet, Jault, Canet, \& Fournier]{gillet2010}
Gillet, N., Jault, D., Canet, E., \& Fournier, A., 2010.
\newblock Fast torsional waves and strong magnetic field within the {E}arth's
  core, {\it Nature\/}, {\bf 465}, 74--77.

\bibitem[Gillet et~al.(2011)Gillet, Schaeffer, \& Jault]{gillet11}
Gillet, N., Schaeffer, N., \& Jault, D., 2011.
\newblock Rationale and geophysical evidence for quasi-geostrophic rapid
  dynamics within the {E}arth's outer core, {\it Phys. Earth Planet. Inter.\/},
  {\bf 187}, 380--390.

\bibitem[Gillet et~al.(2015)Gillet, Jault, \& Finlay]{gillet2015}
Gillet, N., Jault, D., \& Finlay, C.~C., 2015.
\newblock Planetary gyre, time-dependent eddies, torsional waves, and
  equatorial jets at the {E}arth's core surface, {\it J. Geophys. Res. Solid
  Earth\/}, {\bf 120}, 3991--4013.

\bibitem[Gillet et~al.(2017)Gillet, Jault, \& Canet]{gillet2017}
Gillet, N., Jault, D., \& Canet, E., 2017.
\newblock Excitation of traveling torsional normal modes in an {E}arth's core
  model, {\it Geophys. J. Int.\/}, {\bf 210}, 1503--1516.

\bibitem[Gross(2015)]{Gross2015}
Gross, R., 2015.
\newblock Earth rotation variations – long period, in {\em Treatise on
  Geophysics\/}, chap. 3.09, pp. 215--261, ed. Schubert, G., Elsevier, Oxford.

\bibitem[Guervilly \& Cardin(2016)]{guervilly16}
Guervilly, C. \& Cardin, P., 2016.
\newblock Subcritical convection of liquid metals in a rotating sphere using a
  quasi-geostrophic model, {\it J. Fluid Mech.\/}, {\bf 808}, 61--89.

\bibitem[Guervilly et~al.(2014)Guervilly, Hughes, \& Jones]{guervilly2014}
Guervilly, C., Hughes, D.~W., \& Jones, C.~A., 2014.
\newblock Large-scale vortices in rapidly rotating {R}ayleigh--{B}\'enard
  convection, {\it J. Fluid Mech.\/}, {\bf 758}, 407--435.

\bibitem[He \& Sun(2007)]{He+2007}
He, Y. \& Sun, W., 2007.
\newblock Stability and convergence of the crank-nicolson/adams-bashforth
  scheme for the time-dependent navier-stokes equations, {\it SIAM J. Numer.
  Anal.\/}, {\bf 45}(2), 837--869.

\bibitem[Holme(2015)]{Holme2015}
Holme, R., 2015.
\newblock Large-scale flow in the core, in {\em Treatise on Geophysics\/},
  chap. 8.04, pp. 91--113, ed. Schubert, G., Elsevier, Oxford.

\bibitem[Holme \& de~Viron(2013)]{Holme2013}
Holme, R. \& de~Viron, O., 2013.
\newblock Characterization and implications of intradecadal variations in
  length of day, {\it Nature\/}, {\bf 499}, 202--205.

\bibitem[Hough(1897)]{hough1897}
Hough, S.~S., 1897.
\newblock On the {A}pplication of {H}armonic {A}nalysis to the {D}ynamical
  {T}heory of the {T}ides. {P}art {I}. {O}n {L}aplace's `{O}scillations of the
  {F}irst {S}pecies,' and on the {D}ynamics of {O}cean {C}urrents, {\it Philos.
  Trans. R. Soc. London, Ser. A\/}, {\bf 189}, 201--257.

\bibitem[Hunter(2007)]{Hunter2007}
Hunter, J., 2007.
\newblock Matplotlib: A 2d graphics environment, {\it Comput. Sci. Eng.\/},
  {\bf 9}, 90--95.

\bibitem[Jackson et~al.(1993)Jackson, Bloxham, \& Gubbins]{Jackson1993}
Jackson, A., Bloxham, J., \& Gubbins, D., 1993.
\newblock Time-dependent flow at the core surface and conservation of angular
  momentum in the coupled core-mantle system, in {\em Dynamics of the {E}arth's
  deep interior and {E}arth rotation\/}, vol.~72, pp. 97--107, eds Le~Mou\"el,
  J.-L., Smylie, D.~E., \& Herring, T., {AGU} {G}eophysical {M}onograph,
  Washington, DC.

\bibitem[Jault(2008)]{jault08}
Jault, D., 2008.
\newblock Axial invariance of rapidly varying diffusionless motions in the
  {E}arth's core interior, {\it Phys. Earth Planet. Inter.\/}, {\bf 166},
  67--76.

\bibitem[Jault et~al.(1988)Jault, Gire, \& Le~Mou{\"e}l]{Jault1988}
Jault, D., Gire, C., \& Le~Mou{\"e}l, J.-L., 1988.
\newblock Westward drift, core motions and exchanges of angular momentum
  between core and mantle, {\it Nature\/}, {\bf 333}, 353--356.

\bibitem[Jaupart \& Buffett(2017)]{jaupart17}
Jaupart, E. \& Buffett, B., 2017.
\newblock Generation of {MAC} waves by convection in {E}arth's core, {\it
  Geophys. J. Int.\/}, {\bf 209}, 1326--1336.

\bibitem[Labb\'{e} et~al.(2015)Labb\'{e}, Jault, \& Gillet]{Labbe2015}
Labb\'{e}, F., Jault, D., \& Gillet, N., 2015.
\newblock On magnetostrophic inertia-less waves in quasi-geostrophic models of
  planetary cores, {\it Geophys. Astrophys. Fluid Dyn.\/}, {\bf 109}, 587--610.

\bibitem[Livermore et~al.(2008)Livermore, Ierley, \& Jackson]{livermore2008}
Livermore, P.~W., Ierley, G., \& Jackson, A., 2008.
\newblock The structure of {T}aylor's constraint in three dimensions, {\it
  Proc. R. Soc. London, Ser. A\/}, {\bf 464}, 3149--3174.

\bibitem[Matsui et~al.(2014)Matsui, King, \& Buffett]{Matsui+2014}
Matsui, H., King, E., \& Buffett, B., 2014.
\newblock Multiscale convection in a geodynamo simulation with uniform heat
  flux along the outer boundary, {\it Geochem. Geophys. Geosyst.\/}, {\bf 15},
  3212--3225.

\bibitem[Pais \& Jault(2008)]{pais08}
Pais, M.~A. \& Jault, D., 2008.
\newblock Quasi-geostrophic flows responsible for the secular variation of the
  {E}arth's magnetic field, {\it Geophys. J. Int.\/}, {\bf 173}, 421--443.

\bibitem[Proudman(1916)]{proudman1916}
Proudman, J., 1916.
\newblock On the motion of solids in a liquid possessing vorticity, {\it Proc.
  R. Soc. Lond., A\/}, {\bf 92}, 408--424.

\bibitem[Roberts(2015)]{Roberts2015}
Roberts, P., 2015.
\newblock Theory of the geodynamo, in {\em Treatise on Geophysics\/}, chap.
  8.03, pp. 57--90, ed. Schubert, G., Elsevier, Oxford.

\bibitem[Roberts \& King(2013)]{roberts13}
Roberts, P. \& King, E., 2013.
\newblock On the genesis of the {E}arth's magnetism, {\it Rep. Prog. Phys.\/},
  {\bf 76}, 096801.

\bibitem[Roberts et~al.(2007)Roberts, Yu, \& Russell]{Roberts2007}
Roberts, P.~H., Yu, Z., \& Russell, C., 2007.
\newblock On the 60-year signal from the core, {\it Geophys. Astrophys. Fluid
  Dyn.\/}, {\bf 101}, 11--35.

\bibitem[Rotvig \& Jones(2002)]{rotvig2002}
Rotvig, J. \& Jones, C.~A., 2002.
\newblock Rotating convection-driven dynamos at low {E}kman number, {\it Phys.
  Rev. E\/}, {\bf 66}, 056308.

\bibitem[Schaeffer \& Cardin(2005)]{Schaeffer2005}
Schaeffer, N. \& Cardin, P., 2005.
\newblock Quasigeostrophic model of the instabilities of the {S}tewartson layer
  in flat and depth-varying containers, {\it Phys. Fluids\/}, {\bf 17}, 104111.

\bibitem[Schaeffer \& Cardin(2006)]{schaeffer06}
Schaeffer, N. \& Cardin, P., 2006.
\newblock Quasi-geostrophic kinematic dynamos at low magnetic {P}randtl number,
  {\it Earth Planet. Sci. Lett.\/}, {\bf 245}, 595--604.

\bibitem[Schaeffer et~al.(2016)Schaeffer, Lora~Silva, \& Pais]{schaeffer16}
Schaeffer, N., Lora~Silva, E., \& Pais, M.~A., 2016.
\newblock Can core flows inferred from geomagnetic field models explain the
  {E}arth's dynamo?, {\it Geophys. J. Int.\/}, {\bf 204}, 568--877.

\bibitem[Schaeffer et~al.(2017)Schaeffer, Jault, Nataf, \&
  Fournier]{Schaeffer+2017}
Schaeffer, N., Jault, D., Nataf, H.-C., \& Fournier, A., 2017.
\newblock Turbulent geodynamo simulations: a leap towards {E}arth's core, {\it
  Geophys. J. Int.\/}, {\bf 211}, 1--29.

\bibitem[Stellmach et~al.(2014)Stellmach, Lischper, Julien, Vasil, Cheng,
  Ribeiro, King, \& Aurnou]{stellmach2014}
Stellmach, S., Lischper, M., Julien, K., Vasil, G., Cheng, J.~S., Ribeiro, A.,
  King, E.~M., \& Aurnou, J.~M., 2014.
\newblock Approaching the asymptotic regime of rapidly rotating convection:
  Boundary layers versus interior dynamics, {\it Phys. Rev. Lett.\/}, {\bf
  113}, 254501.

\bibitem[Takahashi et~al.(2008)Takahashi, Matsushima, \&
  Honkura]{takahashi2008}
Takahashi, F., Matsushima, M., \& Honkura, Y., 2008.
\newblock Scale variability in convection-driven {MHD} dynamos at low {E}kman
  number, {\it Phys. Earth Planet. Inter.\/}, {\bf 167}, 168--178.

\bibitem[Taylor(1917)]{taylor1917}
Taylor, G.~I., 1917.
\newblock Motion of solids in fluids when the flow is not irrotational, {\it
  Proc. R. Soc. London, Ser. A\/}, {\bf 93}, 99--113.

\bibitem[Taylor(1963)]{Taylor1963}
Taylor, J.~B., 1963.
\newblock The magneto-hydrodynamics of a rotating fluid and the earth's dynamo
  problem, {\it Proc. R. Soc. London, Ser. A\/}, {\bf 274}, 274--283.

\bibitem[Teed et~al.(2014)Teed, Jones, \& Tobias]{teed2014}
Teed, R., Jones, C.~A., \& Tobias, S., 2014.
\newblock The dynamics and excitation of torsional waves in geodynamo
  simulations, {\it Geophys. J. Int.\/}, {\bf 196}, 724--735.

\bibitem[Teed et~al.(2015)Teed, Jones, \& Tobias]{teed2015}
Teed, R.~J., Jones, C.~A., \& Tobias, S.~M., 2015.
\newblock The transition to {E}arth-like torsional oscillations in
  magnetoconvection simulations, {\it Earth Planet. Sci. Lett.\/}, {\bf 419},
  22--31.

\bibitem[Wicht \& Christensen(2010)]{wicht2010}
Wicht, J. \& Christensen, U.~R., 2010.
\newblock Torsional oscillations in dynamo simulations, {\it Geophys. J.
  Int.\/}, {\bf 181}, 1367--1380.

\bibitem[Wu \& Roberts(2015)]{wu2015}
Wu, C.-C. \& Roberts, P.~H., 2015.
\newblock On magnetostrophic mean-field solutions of the geodynamo equations,
  {\it Geophys. Astrophys. Fluid Dyn.\/}, {\bf 109}, 84--110.

\bibitem[Yadav et~al.(2016)Yadav, Gastine, Christensen, Wolk, \&
  Poppenhaeger]{Yadav+2016}
Yadav, R.~K., Gastine, T., Christensen, U.~R., Wolk, S.~J., \& Poppenhaeger,
  K., 2016.
\newblock Approaching a realistic force balance in geodynamo simulations, {\it
  PNAS\/}, {\bf 113}, 12065--12070.

\end{thebibliography}
\appendix
\numberwithin{equation}{section}

\section{Conservation of Axial Angular Momentum}
\label{appendix:angular_momentum}

The total axial angular momentum $\mathcal{L}_{z}$ of the outer core is contained in the concentric
cylinders rotating with axisymmetric angular velocity \vorO.  The time variation of
$\mathcal{L}_{z}$ is then

\begin{equation}
    \frac{1}{4\pi} \pdt \mathcal{L}_{z} = \integral{s_{1}}{s_{2}}{s^{3} L \pdt \pl \vorO \pr}{s} \, .
    \label{eq:dt_Lz2}
\end{equation}

As shown by \Equation{\ref{eq:vorz_axi}}, $\pdt \pl \vorO \pr$ can be decomposed into four component
torques.  Defining the linear operator $\mathcal{G}(f) = \integral{s_{1}}{s_{2}}{s^{3} L f}{s}$,
\Equation{\ref{eq:dt_Lz2}} may be rewritten as a sum:

\begin{align}
    \label{eq:dt_Lz_3}
    \frac{1}{4\pi} \pdt \mathcal{L}_{z}
        &= \mathcal{G} \pl \gammaV \pr
         + \mathcal{G} \pl \Gamma_{\mathrm{L}_{1}} \pr
         + \mathcal{G} \pl \Gamma_{\mathrm{L}_{2}} \pr
         + \mathcal{G} \pl \gammaR \pr \, .
\end{align}

With the input torques

\begin{align}
    \label{eq:gammaVdef}
    \gammaV    &= \frac{E}{s^{3} L} \pds \pl s^{3} L \pds \pl \vorO \pr \pr \, , \\
    \label{eq:gammaL1def}
    \gammaLone &= \frac{1}{s^{3} L} \pds \pl s^{3} L \backbs \curO \pr \, , \\
    \label{eq:gammaL2def}
    \gammaLtwo &= \frac{1}{s} \pl \moyp{\frac{\pbs}{s} \pds \pl s \pbp \pr} \pr \, , \\
    \label{eq:gammaRdef}
    \gammaR    &= - \frac{1}{s} \pl \moyp{\frac{\pus}{s} \pds \pl s \pup \pr} \pr \, ,
\end{align}

the first two terms on the right-hand side of \Equation{\ref{eq:dt_Lz_3}} reduce to
$\mathcal{G}(\gammaV) = \left. s^{3} L \pds \pl \vorO \pr \right|_{s_{1}}^{s_{2}}$ and
$\mathcal{G}(\gammaLone) = \left. s^{3} L \backbs \curO \right|_{s_{1}}^{s_{2}}$.  Because we assume a stress-free condition on $\axiup$ and $\moyp{\pbp} = 0$ at $s_1$ and $s_2$, both terms are zero, reducing
\Equation{\ref{eq:dt_Lz_3}} to

\begin{align}
    \frac{1}{4\pi} \pdt \mathcal{L}_{z}
        &= \mathcal{G} \pl \Gamma_{\mathrm{L}_{2}} \pr
         + \mathcal{G} \pl \gammaR \pr \, .
\end{align}

Since both \uu\ and \bb\ are solenoidal and periodic in azimuth, \Equations{\ref{eq:gammaL2def}} and
(\ref{eq:gammaRdef}) may be rewritten as

\begin{align}
    \gammaLtwo
    &= \phantom{-}
       \frac{1}{s} \pl \frac{1}{s^{2} L} \pds \pl s^{2} L \, \moyp{\pbs \pbp} \pr
                     - \moyp{\pl \beta \pbs - \pdz \pbz \pr \pbp} \pr \, , \\
    \gammaR
    &= -
       \frac{1}{s} \pl \frac{1}{s^{2} L} \pds \pl s^{2} L \, \moyp{\pus \pup} \pr
                     - \moyp{\pl \beta \pus - \pdz \puz \pr \pup} \pr \, .
\end{align}

Application of the $\mathcal{G}$ operator to the first term on the right-hand side of each equation
again produces terms which depend only on boundary values: $s^{2} L \left.\moyp{\pbs
  \pbp}\right|_{s_{1}}^{s_{2}}$ and $s^{2} L \left.\moyp{\pus \pup}\right|_{s_{1}}^{s_{2}}$.
Because we use no-penetration ($\psi = 0 \, \rightarrow \, \pus = 0$) and $a = 0$ ($\rightarrow \,
\pbs = 0$) boundary conditions, both are zero.  This leaves

\begin{equation}
    \label{eq:dt_Lz4}
    \frac{1}{4 \pi} \pdt \mathcal{L}_{z}
    = \mathcal{G} \pl \moyp{\pl \beta \pus - \pdz \puz \pr \pup} \pr
    - \mathcal{G} \pl \moyp{\pl \beta \pbs - \pdz \pbz \pr \pbp} \pr \, .
\end{equation}

An axial profile must be assumed for both the \uu\ and \bb\ fields.  The no-penetration boundary
condition at the spherical top and bottom boundaries dictates that, defining the boundary's normal
vector as ${\bf\hat{n}}$,

\begin{equation}
    \label{eq:no_penetration}
    \uu \cdot {\bf\hat{n}} = 0 \quad \Rightarrow \quad \puz(L) = - \frac{s}{L} \pus \, .
\end{equation}

Assuming that the $\puz$ velocity profile varies linearly with $z$, and is zero in the equatorial
plane,

\begin{equation}
    \label{eq:duz_profile}
    \pdz\puz = \beta \pus \, ,
\end{equation}

which is consistent with mass conservation ($\del \cdot \uu = 0$), and with the definition of the flow
used in \Equations{\ref{eq:full_field_definitions}} and (\ref{eq:velocity_field_definition}).  This
causes the first term on the right-hand side of \Equation{\ref{eq:dt_Lz4}} to vanish.  Our definition
of the magnetic field perturbation $\bb$ in \Equations{\ref{eq:full_field_definitions}} and
(\ref{eq:magnetic_field_definition}) follows the same form as that of $\uu$:

\begin{equation}
    \label{eq:dbz_profile}
    \pdz\pbz = \beta \pbs \, .
\end{equation}

Substituting \Equations{\ref{eq:duz_profile}} and (\ref{eq:dbz_profile}) into
\Equation{\ref{eq:dt_Lz4}} causes the remaining terms on the right-hand side of Eq.~(\ref{eq:dt_Lz_3})
to vanish.  Thus, the
angular momentum is conserved:

\begin{equation}
    \pdt \mathcal{L}_{z} = 0 \, .
    \label{eq:conservation_of_angular_momentum}
\end{equation}

\end{document}